\documentclass[groupedaddress,aps,pra,superscriptaddress,showpacs,twocolumn,prl]{revtex4}%
\usepackage{epsfig,dsfont,amssymb,amsmath,amsthm,amsfonts,amsbsy,mathrsfs}
\usepackage{graphicx, color}
\usepackage{epstopdf}
\usepackage{mathdots}
\usepackage{float}
\usepackage{xcolor}
\newcommand{\tr}{{\rm Tr}}

\begin{document}

\title{Polygamy relations for tripartite and multipartite quantum systems}

\author{Yanying Liang}
\affiliation{College of Mathematics and Informatics, South China Agricultural University, Guangzhou 510642, China}
\author{Haozhen Situ}
\thanks{Corresponding author: situhaozhen@gmail.com }
\affiliation{College of Mathematics and Informatics, South China Agricultural University, Guangzhou 510642, China}
\author{Zhu-Jun Zheng}
\affiliation{School of Mathematics, South China University of Technology, Guangzhou, 510641, China}
\affiliation{Laboratory of Quantum Science and Engineering, South China University of Technology, Guangzhou, 510642, China}

\begin{abstract}
We study the polygamy property for tripartite and multipartite quantum systems. In tripartite system,
we build a solution set for polygamy in tripartite system and find a lower bound of the set, which can be a sufficient and necessary condition for any quantum entanglement of assistance $Q$ to be polygamous. In multipartite system, we firstly provide generalized definitions for polygamy in two kind of divisions of $n$-qubit systems, and then build polygamy inequalities with a polygamy power $\beta$, repectively. Moreover, we use right triangle and tetrahedron to explain our polygamy relations according to the new definitions.
\end{abstract}
\maketitle

\section{Introduction}

As one of the most impressive phenomena in quantum information theory, quantum entanglement \cite{qe} has been extensively explored in recent years focusing on the monogamy or polygamy law \cite{CKW,Gour2005pra, Osborne, Gour,
Ouyongchengpra2,Zhuxuena2014pra,Bai,Choi,Luo,Kim2009,Kim2009302,Lizongguo,
Streltsov,Kim2012pra,Liusiyuan,Reid,Oliveira2014pra,Regula2014prl,Eltschka,Zhuxuena2015pra,Lancien,Song, Luo2016pra,Jia,
 Kim2016pra,Chengshuming,GG}. Since the monogamy and polygamy are the fundamental properties of quantum entanglement. At the early stage, Coffman, Kundu, and Wotters (CKW) showed the monogamy of an quantum entanglement measure $E$ between finite dimensional Hilbert space $H_A$, $H_A$ and $H_C$ has the following form \cite{CKW}:
\begin{equation}\label{CKW}
E(\rho_{A|BC})\geq E(\rho_{AB})+E(\rho_{AC})
\end{equation}
in which $\rho_{AB}=Tr_C(\rho_{ABC})$ and $\rho_{AC}=Tr_B(\rho_{ABC})$,  $\rho_{ABC}\in H_A\otimes H_B\otimes H_C$. 

Later, researchers found some entanglement measures which does not satisfy the inequality (\ref{CKW}). This may reminds us to consider whether the function we are using to quantify monogamy of entanglement can be modified and what the essence of monogamy actually is.
To address this issue, in 2016, the authors proposed a question that ``should entanglement measures be monogamous or faithful?\cite{Lancien}" They said that $E$ is monogamous if the following relation satisfied
\begin{equation}\label{fmonogamy}
E(\rho_{A|BC})\geq f(E(\rho_{AB}),E(\rho_{AC})),
\end{equation}
for some nontrival continuous function $f$ which satisfies $f(x,y)\geq max\{x,y\}$ with some ranges of $x$ and $y$.
In 2018, Gilad Gour and Yu Guo \cite{GG} showed us a fine-grained definition to quantify the monogamy of entanglement. They states that $E$ is monogamous if $E$ satisfies 
\begin{equation}\label{guoyu1}
E(\rho_{A|BC})= E(\rho_{AB})
\end{equation} 
for any $\rho_{ABC}\in H_A\otimes H_B\otimes H_C$, we have $E(\rho_{AC})=0$. In other words, if $A$
shares the entanglement with system $B$ in maximum amount, then there is no entanglement left to share with $C$. The condition for monogamous given in inequality (\ref{CKW}) is stronger than the one given by \cite{GG}.  
Using this novel definition above, Yu Guo \emph{et al.} have developed many conditions for monogamy in terms of different entanglement measures\cite{GY1,GY2,GY3,GY4,GY5,GY6}.
In 2022, the authors in \cite{jin2022} also presented a monogamy relation of entanglement with equality, 
\begin{equation}\label{jin1}
E(\rho_{A|BC})= E(\rho_{AB})+\mu E(\rho_{AC}),
\end{equation}
here $\mu>0$ and $E(\rho_{AB})\geq E(\rho_{AC})$. Comparing these two definitions in equality, we find that equality(\ref{jin1}) is stronger than the definition. Indeed, if $E$ is monogamous having the form of (\ref{jin1}) with $E(\rho_{AB})\geq E(\rho_{AC})$, and if $E$ satisfies equality(\ref{guoyu1}), we easily have $E(\rho_{AC})=0$.

Dually, the polygamy relation for the dual concept of entanglement measure is expressed as
\begin{equation}\label{polybase}
Q(\rho_{A|BC})\leq Q(\rho_{AB})+Q(\rho_{AC}).
\end{equation}
If there exists $0<\beta< \infty$ such that 
\begin{equation}\label{polybeta}
Q^\beta(\rho_{A|BC})\leq Q^\beta(\rho_{AB})+Q^\beta(\rho_{AC}).
\end{equation}
then we say $Q$ is called $\beta$-polygamous.
As far as we know, the entanglement of assistance, and the entanglement of assistance in terms of Tsallis-q entropy have polygamy relations in the form of (\ref{polybeta}) in any multipartite systems \cite{Kim2012pra,Kim2016pra,kim2010}. For multiqubits quantum systems, concurrence of assistance, negativity of assistance, and tangle of assistance are $\beta$-polygamous \cite{Gour2005pra,Gour,Kim2009,Lizongguo}. As for other entanglement correlations in multipartite or  higher dimensional systems, there are many polygamy problems waiting for us to do, which comes out the main goal of this paper. 

Recently, there have been several results about polygamy relations.
In 2018, Yu Guo \cite{GYpoly} found an interesting definition of polygamy in the form of inequality, he said that $Q$ is polygamous when
\begin{equation}\label{guopoly}
Q(\rho_{A|BC})> max\{Q(\rho_{AB}),Q(\rho_{AC})\}>0,
\end{equation}
then one gets $ min\{Q(\rho_{AB}),Q(\rho_{AC})\}>0$.
Later in 2023, the authors in \cite{jin2023} delivered an equality for polygamy. They call $Q$ is polygamous if 
\begin{equation}\label{jinpoly}
Q(\rho_{A|BC})= Q(\rho_{AB})+\gamma Q(\rho_{AC}),
\end{equation}
here $\gamma\geq0$ and $Q(\rho_{AB})\geq Q(\rho_{AC})$.
Comparing these two definitions, we find the definition in \cite{jin2023} is stronger than the one for polygamy in \cite{GYpoly}. In fact, for any $\rho_{ABC}\in H_A\otimes H_B\otimes H_C$, if $\rho_{ABC}$ satisfies condition (\ref{jinpoly}) with $\gamma\geq0$ and $Q(\rho_{AB})\geq Q(\rho_{AC})$, then when $Q(\rho_{A|BC})\geq Q(\rho_{AB})$, we have
\begin{equation}
\gamma Q(\rho_{AC})=Q(\rho_{A|BC})- Q(\rho_{AB})>0,
\end{equation}
with $\gamma\geq0$, which leads to $Q(\rho_{AC})>0$, but not vice versa. 
However, both Ref.\cite{GYpoly} and Ref.\cite{jin2023} do not generalize these results into multipartite systems.

Therefore, specifically, we firstly find a solution set with a lower bound for polygamy in tripartite system, which may reflect the difference between monogamy and polygamy. This part is interesting and can be applied from tripartite into multipartite systems in the rest of this paper. Then we give a new generalized definition of polygamy in multipartite system using two different divisions, and derive polygamy inequalities of $\beta$-power using new definitions. Moreover, in the corresponding examples, we use geometry schematic illustration to explain our results.

\section{Polygamy for tripartite systems using a solution set}

For convenient, we denote $Q_{A|BC}=Q(\rho_{A|BC})$, $Q_{AB}=Q(\rho_{AB})$, and $Q_{AC}=Q(\rho_{AC})$.

Considering a quantum state $\rho_{ABC}\in H_A\otimes H_B\otimes H_C$, we propose a state-dependent parameter $k_{\rho_{ABC}}\geq0$ such that
\begin{eqnarray}\label{def}
k_{\rho_{ABC}}\left(Q^\mu_{A|BC}-max\{Q^\mu_{AB},Q^\mu_{AC}\}\right)=min\{Q^\mu_{AB},Q^\mu_{AC}\},
\end{eqnarray}
with $\mu>0$.

Comparing this equality for polygamy with the one defined in \cite{jin2023}, set $\mu=1$ and $k_{\rho_{ABC}}=\frac{1}{\gamma}$, the equality(\ref{def}) becomes relation (\ref{jinpoly}) in \cite{jin2023}.  Comparing this equality(\ref{def}) for polygamy with the one of inequality in \cite{GYpoly}, if $\mu=1$ and $Q(\rho_{A|BC})> max\{Q(\rho_{AB}),Q(\rho_{AC})\}>0$, using the equality(\ref{def}), we have $ min\{Q(\rho_{AB}),Q(\rho_{AC})\}>0$. 

We denote a set of parameters $k_{\rho_{ABC}}$ by
\begin{eqnarray}\label{defset}
&&\mathbf{K_\mu}=\{k_{\rho_{ABC}}|k_{\rho_{ABC}}\left(Q^\mu_{A|BC}-max\{Q^\mu_{AB},Q^\mu_{AC}\}\right)\nonumber\\
&&=min\{Q^\mu_{AB},Q^\mu_{AC}\},\rho_{ABC}\in H_A\otimes H_B\otimes H_C\}
\end{eqnarray}
Next we wish to explore the property of $\mathbf{K_\mu}$ with polygamy nature.

{\bf Theorem 1.} Let $Q$ be a continuous measure of quantum correlation. $Q$ is $\beta$-polygamous if and only if $\mathbf{K_\mu}$ has a lower bound
$M$ with $M\geq1$.

{\sf [Proof]}

\textbf{(1)}
If $Q$ is $\beta$-polygamous, then $Q^\beta(\rho_{A|BC})\leq Q^\beta(\rho_{AB})+Q^\beta(\rho_{AC})$ with some $\beta>0$. As $Q$ is a measure of quantum correlation, it does not increase under partial trace, so $Q_{A|BC}\geq max\{Q_{AB},Q_{AC}\}$. Without loss of generality, we assume $Q_{AB}>Q_{AC}$.

When $Q_{A|BC}> Q_{AB}$, we can take $\mu=\beta$, so $k_{\rho_{ABC}}=\frac{Q^\beta(\rho_{AC})}{Q^\beta(\rho_{A|BC})- Q^\beta(\rho_{AB})}\geq1$ since  $Q$ is $\beta$-polygamous;

When $Q_{A|BC}= Q_{AB}$, then $Q_{AC}\geq0$ using $\beta$-polygamy. The solution for equality(\ref{def}) is all $k_{\rho_{ABC}}\geq0$. 

\textbf{(2)} Without loss of generality, we assume $Q_{AB}>Q_{AC}$.
 Denote $h(\beta)=Q^\beta_{A|BC}-Q^\beta_{AB}-Q^\beta_{AC}$, when $k_{\rho_{ABC}}>0$, we have $Q_{A|BC}>Q_{AB}>Q_{AC}>0$.
Consider
\begin{eqnarray}\label{hbeta}
h(\beta)&=&Q^\beta_{A|BC}-Q^\beta_{AB}-Q^\beta_{AC}\nonumber\\
&=&\left(Q^{\mu_0}_{AB}+\frac{1}{k_{\rho_{ABC}}}Q^{\mu_0}_{AC}\right)^\frac{\beta}{\mu_0}-Q^\beta_{AB}-Q^\beta_{AC}\nonumber\\
&=&Q^\beta_{AB}(1+\frac{1}{k_{\rho_{ABC}}}\frac{Q^{\mu_0}_{AC}}{Q^{\mu_0}_{AB}})^\frac{\beta}{\mu_0}-Q^\beta_{AB}-Q^\beta_{AC}\nonumber\\
&\leq&Q^\beta_{AB}[1+\frac{\beta}{k_{\rho_{ABC}}\mu_0}(\frac{Q_{AC}}{Q_{AB}})^{\mu_0}]-Q^\beta_{AB}-Q^\beta_{AC}\nonumber\\
&=&Q^{\mu_0}  _{AC}(\frac{\beta}{k_{\rho_{ABC}}\mu_0}Q^{\beta-\mu_0}_{AB}-Q^{\beta-\mu_0}_{AC}),
\end{eqnarray}
where the inequality is due to $(1+t)^x\leq 1+xt$ with $0\leq x\leq 1$ and $t>0$. Since $k_{\rho_{ABC}}\geq M$ and $M\geq1$, then we can choose $\beta> M\mu_0$, we have $h(\beta)\leq 0$.

When $k_{\rho_{ABC}}=0$, we easily get $Q_{AC}=0$. Since $Q_{A|BC}\geq max\{Q_{AB},Q_{AC}\}$, then $Q^\beta_{A|BC}\leq Q^\beta_{AB}+Q^\beta_{AC}$ holds if and only if $Q_{A|BC}=Q_{AB}$.

\qed

From the proof above, we can conclude that if $Q$ is $x$-polygamous, then $Q$ is also $y$-polygamous with $x\geq y$.

In Theorem 1 of Ref.\cite{zhu2023}, the authors give dual results for monogamy: $E$ is monogamous with 
$E^\alpha(\rho_{A|BC})\geq E^\alpha(\rho_{AB})+E^\alpha(\rho_{AC})$ if and only if the set of parameters $k_{\rho_{ABC}}$ is a bounded set, which is quite different from our results.  This may reflect that the difference between monogamy and polygamy relations is determined by the set of $\mathbf{K_\mu}$.

To illustrate better, we give two examples using concurrence of assistance in three-qubit systems. 
Consider the three-qubit state 
\begin{eqnarray}\label{example1}
|\psi\rangle&=&\lambda_0|000\rangle+\lambda_1e^{i{\varphi}}|100\rangle+\lambda_2|101\rangle \nonumber\\
&&+\lambda_3|110\rangle+\lambda_4|111\rangle,
\end{eqnarray}
here $\lambda_i\geq0$, $i=0,1,2,3,4$ and $\sum\limits_{i=0}\limits^4\lambda_i^2=1.$ 

Recall that for a bipartite state $\rho_{AB}$, the concurrence of assistance is defined as 
$C_a(\rho_{AB})=\max\limits_{\{p_i,|\psi_i\rangle_{AB}\}}\sum_ip_iC(|\psi_i\rangle_{AB})$,
where the maximum is taken over all possible pure state decompositions of $\rho_{AB}=\sum\limits_{i}p_i|\psi_i\rangle_{AB}\langle\psi_i|.$
For any pure states $|\psi\rangle_{AB}$, we have $C(|\psi\rangle_{AB})=\sqrt{{2\left[1-\mathrm{Tr}(\rho_A^2)\right]}}$, and $\rho_A=\mathrm{Tr}_B(|\psi\rangle_{AB}\langle\psi|)$.
Since for any pure states $\rho_{AB}=|\psi\rangle_{AB}\langle\psi|$, one has $C_a(\rho_{AB})=C(|\psi\rangle_{AB})$,
then 
$C_a(\rho_{A|BC})=2\lambda_0\sqrt{{\lambda_2^2+\lambda_3^2+\lambda_4^2}},$
$C_a(\rho_{AB})=2\lambda_0\sqrt{{\lambda_2^2+\lambda_4^2}}$ and $C_a(\rho_{AC})=2\lambda_0\sqrt{{\lambda_3^2+\lambda_4^2}}$.
For convenient, assume $\lambda_2 < \lambda_3$, then equality(\ref{def}) with $\mu=1$ becomes 
$k_{|\psi\rangle_{ABC}}(\lambda_0\sqrt{\lambda^2_2+\lambda^2_3+\lambda^2_4}-\lambda_0\sqrt{\lambda^2_3+\lambda^2_4})=\lambda_0\sqrt{\lambda^2_2+\lambda^2_4}$, 
with the following solutions,
\begin{equation}\label{x11}
k_{|\psi\rangle_{ABC}}=\left\{
\begin{aligned}
&~~~ 0,~~~ ~~~~~~ \lambda_0=0;\\
&~~~ 0,~~~~~~~~~  \lambda_0\not=0~~and~~ \lambda_2=0;\\
&\sqrt{1+\frac{\lambda^2_4}{\lambda^2_2}},~~~\lambda_0\not=0~~and~~ \lambda_2\not=0.
\end{aligned}
\right.
\end{equation}
Furthermore, we get $k_{|\psi\rangle_{ABC}}\geq \sqrt{2}+1$ when $\lambda_0\not=0~~and~~ \lambda_2\not=0.$ So taking $M=\sqrt{2}+1$, we have $C_a(\rho_{A|BC})\leq (\sqrt{2}-1)\min\{C_a(\rho_{AB}), C_a(\rho_{AB})\}+\max\{C_a(\rho_{AB}), C_a(\rho_{AB})\}$.

\section{New definition of polygamy for multipartite systems}

{\bf Definition 1.} Let $Q$ be a continuous measure of quantum correlation. $Q$ is called polygamous for any state $\rho_{A_1A_2\cdots A_n}\in H_{A_1}\otimes H_{A_2}\cdots\otimes H_{A_n}$
if it satisfies
\begin{eqnarray}\label{defmulti}
Q_{A_1|A_2\cdots A_n}>max\{Q_{A_1A_2},Q_{A_1A_3},\cdots,Q_{A_1A_n}\}>0,
\end{eqnarray}
then $min\{Q_{A_1A_2},Q_{A_1A_3},\cdots,Q_{A_1A_n}\}>0$.

The definition of polygamy generalizes the one of tripartite systems in Ref.\cite{GYpoly} into multipartite quantum systems. In the following, we will develop more generalized results for polygamy using this new definition, which is meaningful for us to learn the essence of polygamy in multipartite system.

In Ref.\cite{Kim2012pra}, using entanglement of assistance, the author proves a polygamy inequality of multiparty entanglement,
\begin{equation}\label{polyeoa}
E_a(\rho_{A_1|A_2\cdots A_n})\leq E_a(\rho_{A_1A_2})+E_a(\rho_{A_1A_3})+\cdots+E_a(\rho_{A_1A_n}),
\end{equation}
which not only gives a lower bound of how much entanglement can be created
on bipartite subsystems with assistance of the other parties, but provides an upper
bound on the shareability of bipartite entanglement in multiparty quantum systems. Here using the definition above, we can also derive a similar inequality.

{\bf Theorem 2.} Let $Q$ be a continuous measure of quantum correlation. $Q$ is polygamous according to Definition 1
if and only if there exists $\beta>0$ such that 
\begin{equation}\label{polymulti}
Q^\beta(\rho_{A_1|A_2\cdots A_n})\leq Q^\beta(\rho_{A_1A_2})+Q^\beta(\rho_{A_1A_3})+\cdots+Q^\beta(\rho_{A_1A_n}),
\end{equation}
for any state $\rho_{A_1A_2\cdots A_n}\in H_{A_1}\otimes H_{A_2}\cdots\otimes H_{A_n}$.

{\sf [Proof]}
For any state $\rho_{A_1A_2\cdots A_n}\in H_{A_1}\otimes H_{A_2}\cdots\otimes H_{A_n}$, we denote that $Q(\rho_{A_1|A_2\cdots A_n})=x$,$Q(\rho_{A_1A_2})=y_1$,$Q(\rho_{A_1A_3})=y_2$,$\cdots$,$Q(\rho_{A_1A_n})=y_{n-1}$, so next we need to show $x^\beta\leq y_1^\beta+y_2^\beta+\cdots+y_{n-1}^\beta.$

If $x<y_i$ for some $i=1,2,\cdots,{n-1}$, the conclusion is obvious. If $x>max\{y_i\}>0$ for all $i=1,2,\cdots,{n-1}$, since $Q$ is polygamous according to Definition 1, then $y_i>0$ for all $i=1,2,\cdots,{n-1}$.
Considering all $\frac{y_i}{x}\in (0,1)$, we conclude that there always exists $\gamma>0$ such that 
\begin{equation}\label{xyi}
1\leq (\frac{y_1}{x})^\gamma+(\frac{y_2}{x})^\gamma+\cdots+(\frac{y_{n-1}}{x})^\gamma.
\end{equation}
In fact, we can define 
\begin{equation}\label{xy1}
f(\rho_{A_1|A_2\cdots A_n}):=\sup\limits_\gamma\{\gamma|(\frac{y_1}{x})^\gamma+(\frac{y_2}{x})^\gamma+\cdots+(\frac{y_{n-1}}{x})^\gamma\geq 1\}
\end{equation}
Denote $S(H_{A_1A_2\cdots A_n})\equiv S_{A_1A_2\cdots A_n}$ as the set of density matrices acting on a multipartite Hilbert space $H_{A_1A_2\cdots A_n}$. Owing to the compactness of $S_{A_1A_2\cdots A_n}$ and the continuous of $f$, there always exists some state $\rho_i$ satisfies $\min \limits_{\rho \in S_{A_1A_2\cdots A_n}} (\frac{y_i}{x})^\gamma(\rho)=(\frac{y_i}{x})^\gamma(\rho_i)\rightarrow1$ when $\gamma$ decreases for $i=1,2,\cdots,{n-1}$.
Then we derive from the left of Eq.(\ref{xyi}), 
\begin{equation}\label{xy2}
\sum_i(\frac{y_i}{x})^\gamma(\rho)
\geq \sum_i\min\limits_{\rho \in S_{A_1A_2\cdots A_n}}(\frac{y_i}{x})^\gamma(\rho)
= \sum_i(\frac{y_i}{x})^\gamma(\rho_i)
\geq1
\end{equation}
with a sufficiently small positive $\gamma$ independent of $\rho$ decreases.

Moreover, the $\inf\limits_{\rho_{A_1|A_2\cdots A_n} \in S_{A_1A_2\cdots A_n}}f(\rho_{A_1A_2\cdots A_n})$ in Eq.(\ref{xy1}) can also not be infinity due to the compactness of $S_{A_1A_2\cdots A_n}$ and the continuous of $f$.

\qed

One may notice that for any $\alpha\in [0,\beta]$, from the definition of $f(\rho_{A_1|A_2\cdots A_n})$,  
$$
Q^\beta(\rho_{A_1A_2\cdots A_n})\leq Q^\beta(\rho_{A_1A_2})+Q^\beta(\rho_{A_1A_3})+\cdots+Q^\beta(\rho_{A_1A_n}),
$$
implies
$$
Q^\alpha(\rho_{A_1|A_2\cdots A_n})\leq Q^\alpha(\rho_{A_1A_2})+Q^\beta(\rho_{A_1A_3})+\cdots+Q^\alpha(\rho_{A_1A_n}),
$$
with a polygamy measure $Q$ defined in Definition 1. So we call $\beta(Q)$ as the \textit{polygamy power}  of $Q$, the supremum for $Q$ in Eq.(\ref{polymulti}) for any state $\rho_{A_1A_2\cdots A_n}\in H_{A_1}\otimes H_{A_2}\cdots\otimes H_{A_n}$.

When $Q$ is the entanglement of assistance in multipartite system, it is also continuous, so $\beta(Q)$ also exists in this case. We find several examples of entanglement of assistance in Table I. For a two-qubit pure state $|\psi\rangle_{AB}$, or any Schmidt-rank 2 bipartite quantum state, its Schmidt decomposition is 
\begin{equation}\label{schmidt}
|\psi\rangle_{AB}=\sqrt{\lambda_1}|x_1\rangle_{A}\otimes|y_1\rangle_{B}+\sqrt{\lambda_2}|x_2\rangle_{A}\otimes|y_2\rangle_{B},
\end{equation}
The tangle for pure state $|\psi\rangle_{ABC}$ is defined as \cite{Kim2016pra} 
\begin{equation}\label{tau}
\tau(|\psi\rangle_{ABC})=4det\rho_A,
\end{equation}
while the tangle of assistance is
\begin{equation}\label{taua}
\tau^a(\rho_{AB})=\max_{\{p_k,|\psi_k\rangle\}}\sum_k p_k \tau(|\psi_k\rangle_{AB}),
\end{equation}
in which $\rho_{AB}=\tr|\psi_k\rangle_{ABC}\langle\psi|$ with the maximum taken over all pure-state decompositions of $\rho_{AB}$.
Now for the state in (\ref{schmidt}), we have
\begin{equation}\label{tauandt2}
\tau(|\psi\rangle_{AB})=4\lambda_1\lambda_2=T_2(|\psi\rangle_{AB}),
\end{equation}
where $T_2(|\psi\rangle_{AB})$ is the Tsallis-$q$ entanglement with $q=2$ and $T_q(|\psi\rangle_{AB})=S_q(\rho_A)=\frac{1-\tr(\rho^q)}{q-1}$ with $q>0$ and $q\neq0$. So in Table I, the polygamy relations using $\tau_a$ can be covered by $T_2^a$ with $\beta=1$. Moreover, since $\lim\limits_{q\rightarrow1}T_q^a(\rho_{AB})={E_f}_a(\rho_{AB})$, so in Table I, the polygamy relations using ${E_f}_a$ can be covered by $T_1^a$ with $\beta=1$.

\begin{table}
	\caption{\label{tab:table}The comparison of the polygamy power of several entanglement of assistance in multipartite quantum system: concurrence of assistance $C_a$, negativity of assistance $N_a$, entanglement of assistance ${E_f}_a$ associated with the entanglement of formation $E_f$, tangle of assistance $\tau_a$, ${T_q}^a$ associated with the Tsallis-$q$ entropy measure $T_q$ and ${R_\alpha^a}$ associated with the Rényi-$\alpha$ entropy when $\frac{\sqrt{7}-1}{2}\leq \alpha \leq\frac{\sqrt{13}-1}{2}$.}
	\begin{ruledtabular}
		\begin{tabular}{cccc}
			$Q$& $\beta(Q)$ & System & Reference\\ \colrule
			$C_a$ & $2$& $2^{\otimes n}$ & \cite{Gour2005pra}(Conj. 4)\\
			$\tau_a$& $\geq 1$ & $2^{\otimes n}$ &\cite{Gour}(Conj. 2)\\
			${N}_a$& $\geq 2$ & $2^{\otimes n}$ & \cite{Kim2009}(Thm. 1,pure states)\\
	${{E_f}_a}$&$\geq 1$ & $2^{\otimes n}$ & \cite{Kim2012pra}( Eq.(16))\\
		${T_q^a}$, $q\ge1$ &   $\geq 1$    & any system & \cite{Kim2016pra}(Eq.(41))\\
            ${R_\alpha^a}$ &   $\geq 1$    & any system & \cite{yanying}(Cor. 2 and Thm. 6)\\%
		\end{tabular}
	\end{ruledtabular}
\end{table}

\section{Polygamy for multipartite systems}

For any given bipartite entanglement measure $E$, the entanglement of assistance is defined as \cite{GYpoly} 
\begin{equation}\label{ea}
E_a(\rho^{AB})=\max_{\{p_k,|\psi_k\rangle\}}\sum_k p_k E(|\psi_k\rangle),
\end{equation}
where the maximum is taken over all possible pure state decompositions of $\rho_{AB}$.

{\bf Theorem 3.} Let $E_a$ be an entanglement measure of assistance. If $E_a$ is polygamous according to Definition 1
on pure multipartite states in $H_{A_1A_2\cdots A_n}$, then it is also polygamous on mixed states acting on $H_{A_1A_2\cdots A_n}$.

{\sf [Proof]}

Let $\rho_{A_1A_2\cdots A_n}=\sum_k p_k |\psi_k\rangle\langle\psi_k|_{A_1A_2\cdots A_n}$ be an $n$-qubit mixed state with the optimal decomposition such that
\begin{equation}\label{optimal}
E_a(\rho_{A_1|A_2\cdots A_n})=\sum_k p_k E(|\psi_k\rangle_{A_1|A_2\cdots A_n}),
\end{equation}
for all $p_k>0$.
We assume $E_a(\rho_{A_1|A_2\cdots A_n})>max\{E_a(\rho_{A_1A_2}),E_a(\rho_{A_1A_3}),\cdots,E_a(\rho_{A_1A_{n}})\}>0,$
and $\rho_k^{A_1A_j}=Tr_{A_2\cdots A_{j-1}A_{j+1}\cdots A_n}|\psi_k\rangle\langle\psi_k|_{A_1A_2\cdots A_n}$
for $j=2,3,\cdots,n.$ In the following, we need to show $\min\limits_j\{E_a(\rho_{A_1A_j})|j=2,3,\cdots,n\}>0.$

Since any ensemble of $\rho_{A_1A_j}$ can be obtained from $|\psi\rangle_{A_1A_2\cdots A_n}$ by some local operation and classical communication (LOCC) and LOCC can not increase entanglement on average, then we get
\begin{equation}\label{locc}
 E_a(|\psi_k\rangle_{A_1|A_2\cdots A_n})=E(|\psi_k\rangle_{A_1|A_2\cdots A_n})\geq E_a(\rho_{A_1A_j}),
\end{equation}
Besides, based on the assumption and the convex property of $E_a$, we have
\begin{eqnarray}\label{concave}
 E_a(\rho_{A_1|A_2\cdots A_n})&=&\sum_k p_k E(|\psi_k\rangle_{A_1|A_2\cdots A_n})\nonumber\\
&>&\max\limits_j\{E_a(\rho_{A_1A_j})|j=2,3,\cdots,n\}\nonumber\\
&>&\max\limits_j\{\sum_k p_k E_a(\rho_k^{A_1A_j})|j=2,3,\cdots,n\}\nonumber\\
&=&\sum_k p_k E_a(\rho_k^{A_1A_{j_{max}}}),
\end{eqnarray}
Combining Eq.(\ref{locc}) and Eq.(\ref{concave}), we have that there exists some $k_0$ such that 
\begin{equation}\label{usingpure}
 E(|\psi_{k_0}\rangle_{A_1|A_2\cdots A_n})>E_a(\rho_{k_0}^{A_1A_{j_{max}}}).
\end{equation}

If $E_a(\rho_{k_0}^{A_1A_{j_{max}}})>0$, using $E_a$ polygamous on pure state in multipartite states, for fixed $k_0$, we have $\min\limits_j\{E_a(\rho_{k_0}^{A_1A_j})|j=2,3,\cdots,n\}>0$.  Taking $E_a(\rho_{A_1A_{j_{min}}})=\min\limits_j\{E_a(\rho_{A_1A_j})|j=2,3,\cdots,n\},$
 we get
$$E_a(\rho_{A_1A_{j_{min}}})\geq \sum_k p_k E_a(\rho_k^{A_1A_{j_{min}}})>0.$$

If $E_a(\rho_{k_0}^{A_1A_{j_{max}}})=0$, then we have $\rho_{k_0}^{A_1A_{j_{max}}}=\rho_{k_0}^{A_1}\otimes\rho_{k_0}^{A_{j_{max}}}$
 with $\rho_{k_0}^{A_{j_{max}}}$ a pure state. Note that $\rho_{k_0}^{A_1}$ is not pure, otherwise $E(|\psi_{k_0}\rangle_{A_1A_2\cdots A_n})=0$, contradicted with Eq.(\ref{usingpure}). So we can write $|\psi_{k_0}\rangle_{A_1A_2\cdots A_n}\cong|\psi\rangle_{A_1A_2\cdots A_{\Hat{{j_{max}}}}\cdots A_n}\otimes |\psi\rangle_{A_{j_{max}}}$ in which the symbol hat of ${j_{max}}$ means ${j_{max}}$ is removed. Therefore one can have
 \begin{eqnarray}\label{caseis0}
 E_a(\rho_{k_0}^{A_1|A_2\cdots A_{\Hat{{j_{max}}}}\cdots A_n})&=& E_a(|\psi\rangle_{A_1|A_2\cdots A_{\Hat{{j_{max}}}}\cdots A_n})\nonumber\\
&=&E_a(|\psi_{k_0}\rangle_{A_1|A_2\cdots A_n})\nonumber\\
&>&0.
\end{eqnarray}
Now without loss of generality, taking $A=A_1$, $B=A_2\cdots A_{\Hat{{j_{min}}}} \cdots A_{\Hat{{j_{max}}}}\cdots A_n$,$C=A_{{j_{min}}}$, then for the $n-1$-party quantum state $|\psi\rangle_{A_1A_2\cdots A_{\Hat{{j_{max}}}}\cdots A_n}$, Eq.(\ref{caseis0}) can be rewrite as 
$E_a(\rho_{k_0}^{A|BC})>0$ . Since $j_{min} \in {2,3,\cdots,\hat{j_{max}},\cdots,n}$,
Due to the using of LOCC for pure state on $E_a(|\psi_{k_0}\rangle_{A_1|A_2\cdots A_n})$, we get $E_a(\rho_{k_0}^{A|BC})>E_a(\rho_{k_0}^{AB})$ and $E_a(\rho_{k_0}^{A|BC})>E_a(\rho_{k_0}^{AC})$ from Eq.(\ref{caseis0}).
Without loss of generality, we assume $E_a(\rho_{k_0}^{AB})>E_a(\rho_{k_0}^{AC})$, otherwise, repeat the division for $n-2$-party $AB$.
Using Theorem 2 in Ref.\cite{GYpoly} (the polygamy of $E_a$ on tripartite mixed state), if $E_a(\rho_{k_0}^{A|BC})>E_a(\rho_{k_0}^{AB})>0$, we have $E_a(\rho_{k_0}^{AC})>0$, which is equivalent to $E_a(\rho_{k_0}^{A_1A_{{j_{min}}}})>0$.
\qed

{\bf Theorem 4.} For any entanglement measure $E$, entanglement of assistance $E_a$ is always polygamous according to Definition 1.

{\sf [Proof]}
Combining with Theorem 3, we only need to prove $E_a$ is polygamous for multipartite pure state.

Since 
$$E_a(|\psi\rangle_{A_1|A_2\cdots A_n})>E_a(\rho_{A_1A_2}),$$
$$E_a(|\psi\rangle_{A_1|A_2\cdots A_n})>E_a(\rho_{A_1A_3}),$$
$$\cdots$$
$$E_a(|\psi\rangle_{A_1|A_2\cdots A_n})>E_a(\rho_{A_1A_n}),$$
for any $|\psi\rangle_{A_1A_2\cdots A_n}$, in which $\rho_{A_1A_j}$ with $j=2,3,\cdots,n$ is the reduced state of $|\psi\rangle_{A_1A_2\cdots A_n}$.
If $E_a(|\psi\rangle_{A_1|A_2\cdots A_n})>\max\limits_j\{E_a(\rho_{A_1A_j})\}>0,$ then we will show $\min\limits_j\{E_a(\rho_{A_1A_j})\}>0.$

Set
$A=A_1$,$B_1=A_{{j^1_{max}}}$, $C_1=A_2\cdots A_{\Hat{{j^1_{max}}}}\cdots A_n$,
then the $n$-qubit system $A_1A_2\cdots A_n$ can be treated as a tripartite quantum state. We also denote $E_a(\rho_{A_1A_{j^1_{max}}})=\max\limits_j\{E_a(\rho_{A_1A_j})|j=2,\cdots,n\}$. Using Theorem 3 of Ref.\cite{GYpoly}, any entanglement of assistance is polygamous in tripartite system, if $E_a(|\psi\rangle_{A|B_1C_1})>E_a(\rho_{AB_1})>0$, then $E_a(\rho_{AC_1})>0$.
Again, without loss of generality, set $A=A_1$,$B_2=A_{{j^2_{max}}}$, 
$C_2=A_2\cdots A_{\Hat{{j^2_{max}}}}\cdots A_{\Hat{{j^1_{max}}}}\cdots A_n$,
the $n-1$-qubit system $A_1A_2\cdots A_{\Hat{{j^1_{max}}}}\cdots A_n$ can be treated as a tripartite quantum state, and $E_a(\rho_{A_1A_{j^2_{max}}})=\max\limits_j\{E_a(\rho_{A_1A_j})|j=2,\cdots \Hat{{j^1_{max}}},\cdots,n\}$. Using Theorem 3 of Ref.\cite{GYpoly} again, we have $E_a(\rho_{AC_2})>0$. Thus, repeat these operations $n-2$ times, we get $\min\limits_j\{E_a(\rho_{A_1A_j})\}>0.$

\qed

Now let us consider an example of the properties of polygamy in multipartite quantum system.
For the $n$-qubit $W$-class states
\begin{equation}\label{wclass}
|W\rangle_{A_1A_2\cdots A_n}=a_1|1\cdots0\rangle_{A_1A_2\cdots A_n}+\cdots+a_n|0\cdots1\rangle_{A_1A_2\cdots A_n},
\end{equation}
with $\sum^{n}_{i=1}|a_i|^2=1$, then after computing using the concurrence of assistance, we could see that 
$C_{A_1A_2\cdots A_n}=2|a_1|\sqrt{\sum^{n}_{i=2}|a_i|^2}$ and $C_a(\rho_{A_1A_i})=2|a_1||a_i|$. Then obviously $C_a$ is polygamous according to our definition, and the polygamy relation for $C_a$ in Eq.(\ref{polymulti}) is saturated with $\beta=2$.

\begin{figure}
\centering
\includegraphics[width=7cm]{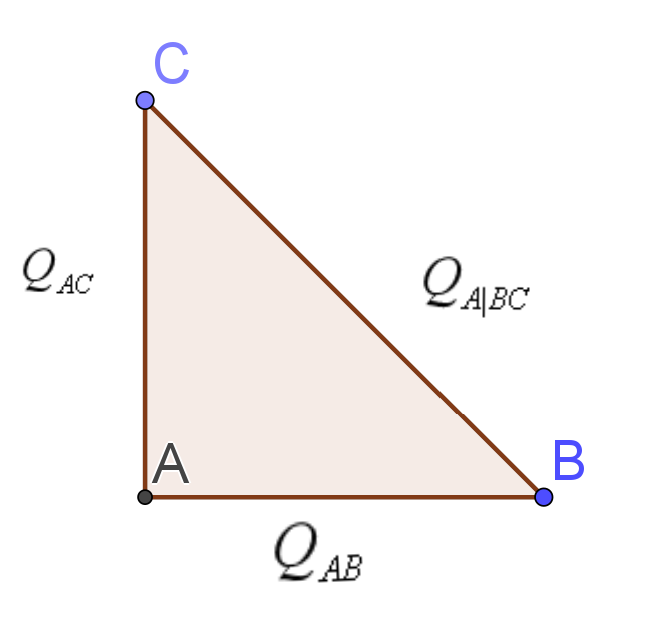}\\
\caption{Polygamy relation according to Definition 1 for $3$-qubit $W$-class states using concurrence.}\label{1}
\end{figure}

\begin{figure}
\centering
\includegraphics[width=9cm]{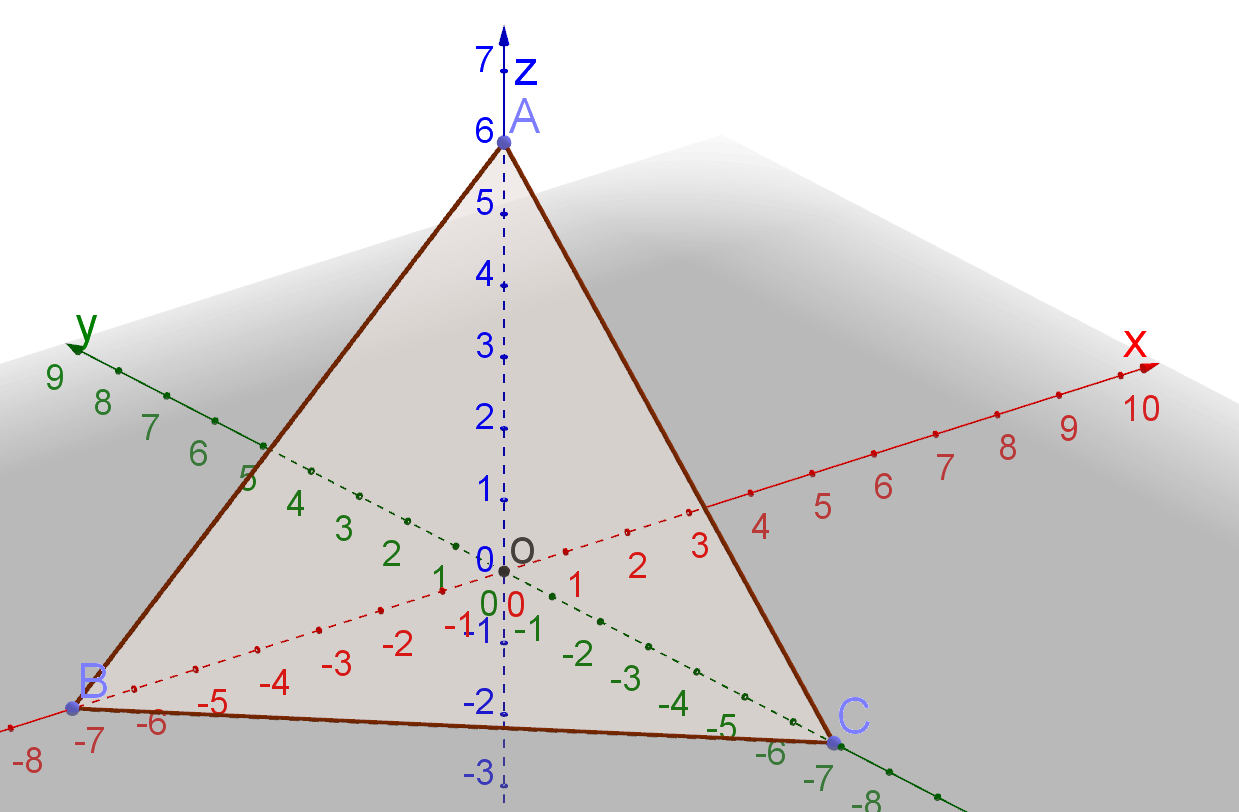}\\
\caption{Polygamy relation according to Definition 1 for  $4$-qubit $W$-class states using concurrence.}\label{2}
\end{figure}

We translate the polygamy relations according to Definition 1 into geometric schematic illustration.
In Figure.(\ref{1}), the right triangle with angle A being $\frac{\pi}{2}$ represents 
\begin{equation}\label{righttriangle}
Q^2_{A|BC}\leq Q^2_{AB}+Q^2_{AC},
\end{equation}
when 
\begin{eqnarray}
Q_{A|BC}>max\{Q_{AB},Q_{AC}\}>0.
\end{eqnarray}
The polygamy relation in Eq.(\ref{righttriangle}) is exactly the Pythagorean theorem, and the condition in Definition 1 is exactly 
the relationship between the right-angled edge and hypotenuse.
Moreover, in $4$-qubit system, we choose the tetrahedron with three faces perpendicular to each other in Figure.(\ref{2}) to explain the polygamy relation
\begin{equation}\label{perpendicular}
Q^2_{A|BCD}=Q^2_{AB}+Q^2_{AC}+Q^2_{AD}.
\end{equation}
Set $S_{ABC}$, $S_{OAB}$, $S_{OAC}$, and $S_{OBC}$ represents $Q_{A|BCD}$, $Q_{AB}$, $Q_{AC}$, and $Q_{AD}$ respectively, in which $S_{ABC}$, $S_{OAB}$, $S_{OAC}$, and $S_{OBC}$ is the area size of triangle $ABC$, $OAB$, $OAC$ and $OBC$. Since we have $S^2_{ABC}=S^2_{OAB}+S^2_{OAC}+S^2_{OBC}$
in geometry theory, then it is exactly the Eq.(\ref{perpendicular}).
The condition in Definition 1 is exactly 
the relationship between these four area sizes.

From the point of a solution set in section 1, we 
consider a quantum state $\rho_{A_1A_2\cdots A_n}\in H_{A_1A_2\cdots A_n}$. Similarly we propose a solution set $\mathbf{K^{n-qubit}_\mu}$ in the following form,
\begin{eqnarray}\label{npartykmu}
\mathbf{K^{n-qubit}_\mu}=\mathbf{K^1_\mu}\cap \mathbf{K^2_\mu} \cap\cdots\cap \mathbf{K^{n-2}_\mu},
\end{eqnarray}
in which
\begin{eqnarray}\label{k1mu}
\mathbf{K^1_\mu}&=&\{k_{\rho_{A_1A_2\cdots A_n}}|k_{\rho_{A_1A_2\cdots A_n}}(Q^\mu_{A_1A_2\cdots A_n}-\nonumber\\
&&max\{Q^\mu_{A_1A_2},Q^\mu_{A_1A_3\cdots A_n}\})\}\nonumber\\
&=&min\{Q^\mu_{A_1A_2},Q^\mu_{A_1A_3\cdots A_n}\},
\end{eqnarray}
\begin{eqnarray}\label{k2mu}
\mathbf{K^2_\mu}&=&\{k_{\rho_{A_1A_3\cdots A_n}}|k_{\rho_{A_1A_3\cdots A_n}}(Q^\mu_{A_1A_3\cdots A_n}-\nonumber\\
&&max\{Q^\mu_{A_1A_3},Q^\mu_{A_1A_4\cdots A_n}\})\}\nonumber\\
&=&min\{Q^\mu_{A_1A_3},Q^\mu_{A_1A_4\cdots A_n}\},
\end{eqnarray}
$$\cdots$$
\begin{eqnarray}\label{kn-1mu}
\mathbf{K^{n-2}_\mu}&=&\{k_{\rho_{A_1A_{n-1}A_n}}|k_{\rho_{A_1A_{n-1}A_n}}(Q^\mu_{A_1A_{n-1}A_n}-\nonumber\\
&&max\{Q^\mu_{A_1A_{n-1}},Q^\mu_{A_1A_n}\})\}\nonumber\\
&=&min\{Q^\mu_{A_1A_{n-1}},Q^\mu_{A_1A_n}\},
\end{eqnarray}
with $\mu>0$.

According to Theorem 1, each $\mathbf{K^{i}_\mu}$ for $i=1,2,\cdots,{n-2}$ has a lower bound $M_i$, therefore set $M=max\{M_i\}$, and $\mathbf{K^{n-qubit}_\mu}$ also has a lower bound $M$.

\section{Polygamy relations between each single qubit and its remaining partners}

Besides the distribution from a 'one-to-group' (such as $A|BC$) entanglement into all 'one-to-one' (such as $AB$, $AC$) entanglement, we want to consider the entanglements between all 'one-to-group' divisions (such as $A|BC$, $B|AC$, $C|AB$). Such a consideration can provide a different view of fundamental entanglement restriction comparing with former sections.
We take the tripartite state $\rho_ABC$ as a beginning.

{\bf Definition 2.} Let $Q$ be a continuous measure of quantum entanglement. $Q$ is called polygamous for any state $\rho_{ABC}$,
when
\begin{eqnarray}\label{onetogroup}
Q_{A|BC}>max\{Q_{B|AC},Q_{C|AB}\}>0,
\end{eqnarray}
we have $min\{Q_{B|AC},Q_{C|AB}\}>0$.

In Ref.\cite{qian}, the author build a set of entanglement inequalities among all $N$ 'one-to-group' marginal bi-partite entanglements. For example, in tripartite system, many entanglement measures, such as concurrence, satisfy the symmetric relation 
\begin{eqnarray}\label{qian}
E_{i|jk} \geq E_{j|ik}+E_{k|ij},
\end{eqnarray}
for $i\neq j\neq k \in \{A,B,C\}$.
In Ref.\cite{jin2023}, the author also present a polygamy relation for $\rho_{ABC}$
\begin{eqnarray}\label{jin}
E_{A|BC} \geq E_{B|AC}+\delta E_{C|AB},
\end{eqnarray}
for some $\delta>0$ and $E_{A|BC} \geq E_{B|AC} \geq E_{C|AB}$.
In fact, Eq.(\ref{qian}) is stronger than Eq.(\ref{jin}) for $\delta=1$; Eq.(\ref{jin}) is stronger than our Eq.(\ref{onetogroup}) for some $\delta$ and the case $E_{A|BC} \geq E_{B|AC} \geq E_{C|AB}$.

{\bf Theorem 5.} Let $Q$ be a continuous measure of quantum correlation. $Q$ is polygamous according to Definition 2
if and only if there exists $\beta>0$ such that 
\begin{equation}\label{polypoweronetogroup}
Q^\beta(\rho_{A|BC})\leq Q^\beta(\rho_{B|AC})+Q^\beta(\rho_{C|AB}),
\end{equation}
for any state $\rho_{ABC}$.

{\sf [Proof]}
For any state $\rho_{ABC}\in H_{A}\otimes H_{B}\cdots\otimes H_{C}$, we denote that $Q(\rho_{A|BC})=x$,$Q(\rho_{B|AC})=y$,$Q(\rho_{C|AB})=z$, so next we need to show $x^\beta\leq y^\beta+z^\beta.$

If $x<y$ or $x<z$, the conclusion is obvious. If $x>max\{y,z\}>0$, since $Q$ is polygamous according to Definition 2, then $y>0$ and $z>0$.
Considering all $\frac{y}{x}\in (0,1)$ and $\frac{z}{x}\in (0,1)$, we conclude that there always exists $\gamma>0$ such that 
\begin{equation}\label{xyz}
1\leq (\frac{y}{x})^\gamma+(\frac{z}{x})^\gamma.
\end{equation}
Since $\gamma$ decreases leads to ${\frac{y}{x}}^\gamma$ and ${\frac{z}{x}}^\gamma$ tending to 1.
Assume 
\begin{equation}\label{fonetogroup}
f(\rho_{A|BC}):=\sup\limits_\gamma\{\gamma|1\leq (\frac{y}{x})^\gamma+(\frac{z}{x})^\gamma\}
\end{equation}
and choose $\beta$ as the greatest lower bound of $f(\rho_{A|BC})$ is enough.
Denote $S(H_{ABC})\equiv S_{ABC}$ as the set of density matrices acting on a tripartite Hilbert space $H_{ABC}$,
Due to $S_{ABC}$ compact and $f$ continuous, $\beta$ can not be infinity.
\qed

Similar to the analysis in Theorem 2, we also call $\beta$ in Theorem 5 \textit{polygamy power}  of $Q$. In other words, any $\sigma \in [0,\beta]$ satisfies the polygamy relations Eq.(\ref{polypoweronetogroup}).
Besides, we note that if the condition in Definition 2 becomes $Q_{B|AC}>max\{Q_{A|BC},Q_{C|AB}\}>0$, or $Q_{C|AB}>max\{Q_{B|AC},Q_{A|BC}\}>0,$, then one can analyze these cases in the similar way and get the similar results.

\section{Conclusion and discussion}

In this paper, we have recalled all the definitions for polygamy and monogamy so far, and compared the advantages of each definition in literature. To find the essence of polygamy based on the former researches, we firstly find a lower bound of a solution set using polygamy relations, which may tells us the difference for polygamy and monogamy through comparing with the state of arts. Then we generalize the definition of polygamy into $n$-qubit systems. To show its efficiency, we build polygamy inequalities with polygamy power, and further prove any entanglement of assistance $E_a$ is polygamous according to our new definition. In the future, we want to explore the polygamy in $n$-qudit or higher dimensional systems to find more results for the nature of quantum entanglement.

\end{document}